\documentclass[aps,prb,twocolumn,a4paper,superscriptaddress,amsmath,amssymb]{revtex4}

\usepackage[dvips]{graphicx}

\renewcommand{\vec}[1]{\mathbf #1} \renewcommand{\i}{\mathrm i}

\newcommand{\pdiff}[2]{\frac{\partial #1}{\partial #2}}
\newcommand{\pdiffdiff}[2]{\frac{\partial^2 #1}{\partial #2^2}}
\renewcommand{\d}{\mathrm d} \renewcommand{\r}{\vec r}
\renewcommand{\Im}[1]{\text{Im}\left\{ #1\right\}}
\renewcommand{\Re}[1]{\text{Re}\left\{ #1\right\}}

\begin{document}
%\twocolumngrid

%Title of paper
\title{Spin transport and magnetoresistance in F/S/F spin valves}

\author{Jan Petter Morten}
\email{jan.morten@phys.ntnu.no}
\author{Arne Brataas}
%\email[]{arne.brataas@phys.ntnu.no}
\affiliation{Department of Physics, Norwegian University of Science
and Technology, 7491 Trondheim, Norway}

\author{Wolfgang Belzig}
%\email{wolfgang.belzig@unibas.ch}
\affiliation{Department of Physics and Astronomy, University of Basel,
Klingelbergstrasse 82, 4056 Basel, Switzerland}

%\date{\today}
\date{24 January 2004}

\begin{abstract}
  We consider spin transport and spin relaxation in superconductors
  using the quasiclassical theory of superconductivity. We include
  spin relaxation due to spin-orbit interaction as well as magnetic
  impurities, and show that the energy dependence of the spin-flip
  rate is different for these two mechanisms. In
  ferromagnet-superconductor-ferromagnet systems made of Co and Al,
  interface resistances can be small compared to bulk resistances.
  This simplifies the description of transport in Co/Al/Co spin
  valves, for which we numerically calculate the temperature and Al
  length dependence of the magnetoresistance.
\end{abstract}

%\keywords{}

\maketitle

\section{\label{sec:intro}Introduction}
%*******************************************************************
Ferromagnetism and superconductivity are two competing phenomena in
condensed matter physics. In conventional low temperature
superconductors, transport of spins beyond the coherence length is
prevented by the formation of spin singlet Cooper pairs.
Consequently, due to the competing ordering of ferromagnets (F) and
superconductors (S) in hybrid structures, many nontrivial physical
effects occur\cite{Beckmann:prl04,Leridon:cond-mat04} and there are
interesting suggested applications such as an absolute spin valve
effect\cite{Huertas-Hernando:prl02} and solid state memory
elements.\cite{fsf}

Experimental studies of F/S contacts in the diffusive limit showed
that the resistance can both decrease and increase relative to
the resistance above the critical temperature ($T_c$) of the
superconductor.\cite{Petrashov:prl99,Giroud:prb98,Aumentado:prb01}
Theoretically it was shown that the temperature dependence of this
resistance depends sensitively on the contact
transparency.\cite{Jedema:prb99,Falko:jetp99,Belzig:prb00} The
resulting resistance is determined by an interplay between the
energy-dependent interface resistance and spin accumulation at the
interfaces due to reduced spin transport into the superconductor.

Transport of spins through the bulk of superconductors was recently
studied experimentally in an F/S/F heterostructure.\cite{Gu:prb02}
Here, a decreased magnetoresistance (MR) in the superconducting state
was interpreted as a loss of spin memory. Theoretical work on bulk spin
transport in superconductors in the inelastic
regime\cite{Yamashita:prb02} and the elastic regime
\cite{Morten:prb04} describes the reduced penetration of spins by spin
flipping and reduced penetration of spin polarized quasiparticles. The
F/S/F system of Ref. \onlinecite{Gu:prb02} has been analyzed by
assuming a spatially homogeneous superconducting order parameter and
neglecting spin flip.\cite{Yamashita:prb03} However, a thorough
understanding of spin transport in F/S systems requires a description
of the spatially dependent order parameter in each component as well as the
quasiparticles driven out of equilibrium. A theoretical description
of the F/S/F system, where the spatial variation of the order
parameter, energy-dependent spin flipping in the superconductor, and
the effect of the interfaces is taken into account, has to the best of
our knowledge not been published.

In order to study the bulk spin transport properties, it is important
to have control over the influence of interfaces. Typically, in spin
valve structures both interface resistances and bulk resistivities
contribute to the MR and are affected by superconductivity. In this
paper, we study a superconducting spin valve system, where the interface
resistances are negligible. In that case, a simplified treatment of the
F/S boundaries is possible so that bulk effects can be studied
independently of interface effects. As we discuss later, a possible
candidate to realize a spin valve with small interface resistance
could be a Co/Al/Co system. To describe the transport through a
superconducting spin valve, we present a theoretical framework that
describes the spin dependent transport in superconductors in linear
response. Spin flip scattering from magnetic impurities as well as
spin-orbit interaction is included in our description, and the full
spatial dependence of the pairing potential is calculated
self-consistently. We use this formalism for numerical calculations of
the magnetization-configuration dependent transport of a Co/Al/Co spin
valve. This demonstrates the suppression of spin transport through the
superconductor.

The paper is organized in the following way: Section \ref{sec:theory}
describes the equations governing elastic transport in a diffusive
superconductor.  Section \ref{sec:model} outlines the specific
geometry studied and the approximations used.  In section
\ref{sec:numerics} we discuss the numerical results.  Section
\ref{sec:conclusion} summarizes and concludes our work.

\section{\label{sec:theory}Transport Theory}
%*********************************************************************
Using the Keldysh theory in the quasiclassical approximation, we have
previously derived kinetic equations for transport of charge, energy,
spin and spin energy in diffusive, $s$-wave superconductors in the
presence of spin-flip scattering by magnetic
impurities.\cite{Morten:prb04} We will
now supplement that treatment with expressions for spin-orbit induced
spin relaxation, and derive the resulting transport equations in linear
response. For an explanation of the notations used below we refer to
Ref.~\onlinecite{Morten:prb04}. 

The spin-orbit interaction Hamiltonian is
\begin{equation}
  \label{eq:hso}
  H_\text{so}=\frac{\gamma}{2}\mathop{\sum}_{\sigma'\sigma}\int\d\r\psi^\dag_{\sigma'}\left\{\left(\boldsymbol{\bar{\tau}}\times\nabla V_\text{imp}\right)_{\sigma'\sigma}\cdot\vec{p}+\text{H.c.}\right\}\psi_\sigma,
\end{equation}
where $\gamma$ is the interaction strength, $\boldsymbol{\bar{\tau}}$
is the vector of Pauli matrices and $V_\text{imp}$ is the impurity
scattering potential. The spin-orbit contribution to the self-energy in
the Eilenberger equation\cite{Schmid:nato81} is
\begin{align}
  \check{\sigma}_\text{so}=-\frac{\i}{2\tau_\text{so}}\frac{1}{4}\hat{\boldsymbol{\alpha}}\hat{\rho}_3\check{g}_s(X,E)\hat{\rho}_3 \hat{\boldsymbol{\alpha}},\label{eq:sosigma}
\end{align}
where we have defined the spin-orbit scattering time
$1/\tau_\text{so}=8\gamma^2p_\text{F}^4/9\tau$. Here $p_F$ is the
Fermi momentum, $\tau$ is the elastic scattering time,
$\boldsymbol{\hat{\alpha}}$ is a vector of 4$\times$4 matrices with
the Pauli matrix and its transpose on the diagonal block, i.e.
$\boldsymbol{\hat{\alpha}}=\text{diag}(\boldsymbol{\bar{\tau}},\boldsymbol{\bar{\tau}}^T)$,
$\hat{\rho_3}=\text{diag}(1,1,-1,-1)$, and $\check{g}_s$ is the
isotropic part of the Green's function in Keldysh-Nambu-spin-space.
Using a convenient representation of the Green's functions, we obtain
equations that determine the distribution functions and currents.

Linearized kinetic equations for charge transport in diffusive
superconductors were obtained by Schmid and
Sch\"{o}n\cite{Schmid:jltp75} and have been successfully applied to
describe various transport phenomena. To study spin-dependent
transport it is necessary to include equations that determine the
spin-current. The relevant equations in the linear response regime are
developed below.  The approximations are valid when deviations from
equilibrium values are small. We also assume that any static
supercurrent is small, i.e. that there is no Josephson effect. The
transport theory is formulated in terms of the physical particle and
energy currents (including particles and holes). These are given by
the distribution functions $h_\text{T}$ and
$h_\text{L}$\cite{Belzig:sm99} and the spin resolved functions
$h_\text{TS}$ and $h_\text{LS}$, as well as generalized diffusion
coefficients $D_\text{T},~D_\text{L}$ and renormalization factors
$\alpha_\text{TT},~\alpha_\text{TSTS}$ for relaxation processes. The
spin-resolved distribution functions can be expressed by the particle
distribution function as
\begin{equation}
  \label{eq:hs}
  h_{\stackrel{\scriptstyle{\text{TS}}}{\scriptstyle{\text{LS}}}}=-\frac{f_\uparrow(E)-f_\downarrow(E)}{2}\mp\frac{f_\uparrow(-E)-f_\downarrow(-E)}{2}.  
\end{equation}
The spectral (retarded) properties depend on the complex function
$\theta(E)$ which is determined by the so-called Usadel
equation.\cite{Belzig:sm99} To describe spin polarized transport in
voltage biased systems in linear response, it unnecessary to calculate
$h_\text{L}$ and $h_\text{LS}$, so the equations that determine these
functions have been omitted below.

The charge current and the spin current in S are given by integrals
over the spectral quantities. The charge and spin current carried by
quasiparticles is\footnote{The quasiparticle charge current depends on
  the coordinate $x$, but conservation of charge is satisfied when
  including the supercurrent contribution so that the total charge
  current does not depend on $x$.}
\begin{align}
  I_\text{charge}^\text{qp}(x)=&\,\frac{1}{2}eAN_0\int_{-\infty}^\infty\d E D_\text{T}(E,x)\pdiff{h_\text{T}}{x},\label{eq:Ncurr}\\
  I_\text{spin}(x)=&\,\frac{1}{2}eAN_0\int_{-\infty}^\infty\d E D_\text{L}(E,x)\pdiff{h_\text{TS}}{x},\label{eq:spincurr}
\end{align}
where $A$ is the area of the wire and $N_0$ is the density of states
at the Fermi level for both spins in the normal state. Additionally,
charge current is carried by the supercurrent, so that the total
charge current is constant. The distribution functions $h_\text{T}$
and $h_\text{TS}$ are determined by the diffusion equations
\begin{align}
  \pdiff{}{x}\left(D_\text{T}\pdiff{h_\text{T}}{x}\right)-2\Delta\alpha_\text{TT}h_\text{T}=&0,\label{eq:linbndhT}\\
  \pdiff{}{x}\left(D_\text{L}\pdiff{h_\text{TS}}{x}\right)-\left(\frac{1}{\tau_\text{m}}\alpha^\text{m}_\text{TSTS}+\frac{1}{\tau_\text{so}}\alpha^\text{so}_\text{TSTS}\right)&h_\text{TS}=0\label{eq:linbndhTS}.
\end{align}
Here $\tau_\text{m}$ is the spin flip scattering time due to magnetic
impurities and $\tau_\text{so}$ the spin flip scattering time due to
spin-orbit coupling, both evaluated in the normal state. The spectral
quantities are given in terms of $\theta(E,x)$. We compute that the
renormalization of the scattering rates are
\begin{subequations}
  \label{eq:alpha}
  \begin{align}
    \alpha_\text{TT}=&\Im{\sinh(\theta)},\\
    \alpha^\text{so}_\text{TSTS}=&\left(\Re{\cosh(\theta)}\right)^2-\left(\Re{\sinh(\theta)}\right)^2,\label{eq:alphaso}\\
    \alpha^\text{m}_\text{TSTS}=&\left(\Re{\cosh(\theta)}\right)^2+\left(\Re{\sinh(\theta)}\right)^2,\label{eq:alphasf}\\
    D_\text{L}=&D\left[\left(\Re{\cosh(\theta)}\right)^2-\left(\Re{\sinh(\theta)}\right)^2\right],\\
    D_\text{T}=&D\left[\left(\Re{\cosh(\theta)}\right)^2+\left(\Im{\sinh(\theta)}\right)^2\right].
  \end{align}
\end{subequations}
The effect of spin-flip scattering by spin-orbit interaction with
renormalization factor $\alpha_\text{TSTS}^\text{so}$ is a new result
that did not appear in our previous article.\cite{Morten:prb04} Its
renormalization is different from the renormalization of the spin-flip
scattering by magnetic impurities. The complex function $\theta$ is
determined by the Usadel equation
\begin{align}
  \hbar D \pdiffdiff{\theta}{x}=-2\i\Delta\cosh(\theta)-2\i E\sinh(\theta)+\frac{3}{4}\frac{\hbar}{\tau_\text{m}}\sinh(2\theta).\label{eq:retardedusadel}
\end{align}
Note that the spin flip term in \eqref{eq:retardedusadel} arises from
magnetic impurities only since spin-orbit scattering does not lead to
pair breaking and consequently does not influence the spectral
properties of the superconductor. This equation must be solved in
conjunction with the self-consistency relation
\begin{align}
  \Delta=\frac{1}{2}N_0\lambda\int_0^{E_\text{D}}\d E~\Re{\sinh\left(\theta\right)}\tanh\left(\frac{\beta E}{2}\right),
  %\Delta=\frac{1}{2}N_0\lambda\int_0^{E_\text{D}}\d E \sinh\left(\Re{\theta}\right)\cos\left(\Im{\theta}\right)\tanh\left(\frac{\beta E}{2}\right),
  \label{eq:selfconsistency}
\end{align}
where $\lambda$ is the electron-electron interaction strength and
$E_\text{D}$ the Debye cut-off energy.

An applied voltage is taken into account as a boundary condition for
the distribution functions, $h_\text{T}$ and $h_\text{TS}$. In a
reservoir with electrochemical potential $\mu$ we have in linear
response the equilibrium distributions
$h^0_\text{T}=-\beta\mu/\left(2\cosh^2(\beta E/2)\right)$ and
$h^0_\text{TS}=0$.

The different renormalization factors $\alpha^\text{so}_\text{TSTS}$
and $\alpha^\text{m}_\text{TSTS}$ arise from spin flipping by
spin-orbit interaction or magnetic impurities. In general,
$\alpha^\text{so}_\text{TSTS}$ and $\alpha^\text{m}_\text{TSTS}$
depend on the spectral properties of the superconductor through
$\theta$. In the BCS limit, valid for large bulk superconductors, the
energy dependence of these factors is completely different and
correspond to the so-called type-I or type-II coherence
factors.\cite{Tinkham,Yafet:pl83} Using the BCS solution for the
Green's functions we find that for energies below the gap (for which
there are no quasiparticles in the BCS limit), both
$\alpha_\text{TSTS}^\text{so}$ and $\alpha_\text{TSTS}^\text{m}$
vanish, and above the gap $\alpha^\text{so}_\text{TSTS}=1$ and
$\alpha^\text{m}_\text{TSTS}=(E^2+\Delta^2)/(E^2-\Delta^2)>1$.
Furthermore, we see from Equations \eqref{eq:alphaso} and
\eqref{eq:alphasf} that for any $\theta(E)$,
$\alpha^\text{so}_\text{TSTS}<\alpha^\text{m}_\text{TSTS}$. This
implies that for a given normal state spin-flip length, the rate of
spin flipping in the superconducting state is higher when the dominant
spin-flip scattering mechanism is caused by magnetic impurities than
if it is caused by spin-orbit interaction.

\section{\label{sec:model}Model}
\begin{figure}[htbp]
\includegraphics[scale=0.35]{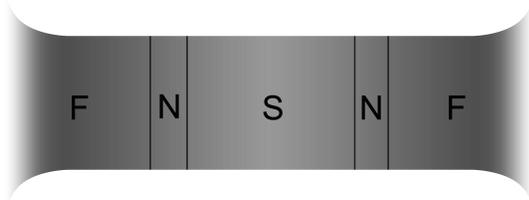}
\caption{Diffusive wire consisting of F,N and S elements.}
\label{fig:pillar}
\end{figure}

%*********************************************************************
We consider an F/N/S/N/F wire as shown in Figure \ref{fig:pillar}. It
is assumed that the ferromagnets (Co) are connected via normal metals
(Cu) to the superconductor (Al). The distribution functions in the
ferromagnet and the normal metal are determined by the Valet-Fert
transport theory\cite{Valet:prb93} and in the superconductor by the
theory described in the previous section. An applied bias causes spin
polarized quasiparticles to be injected into the S layer. We assume
that the magnetization of the F parts are either parallel (P) or
antiparallel (AP). Because of renormalized spin flip rates and a
reduction of the generalized spin diffusion coefficient ($D_\text{L}$)
in the superconductor, the magnetoresistance
$\text{MR}\equiv(R^\text{AP}-R^\text{P})/R^\text{P}$ is reduced for
temperatures below $T_c$ compared to the normal metal state.

In order to determine the dominant contributions to the resistance of
the system, we examine the magnitude of the resistance of the F/N
interface ($R_\text{F/N}$) compared to the bulk resistance in F within
a spin-flip length ($R^\text{F}_\text{sf}$). The latter quantity is
the largest resistance of the ferromagnet within its spin-active part.
To this end, consider the ratio
\begin{equation}
  \frac{R_\text{F/N}}{R_\text{sf}^\text{F}}=\frac{AR_\text{F/N}}{\rho^\text{F}l^\text{F}_\text{sf}}.\label{eq:RfnRsf}
\end{equation}
We assume that F layers are made of Co, N layers of Cu and the S layer
of Al. The bulk resistance of Cu as well as the proximity effect is
neglected since the Cu layer is very thin, and in addition the typical
resistivity of Cu is smaller than that of Co or Al. The interface
resistance ($AR_\text{F/N}$), resistivity ($\rho^\text{F}$) and spin
diffusion length ($l^\text{F}_\text{sf}$) for Co is reviewed in Ref.
\onlinecite{Bass:jmmm99}. It is found that $AR_\text{Cu/Co}\sim
0.5~\text{f}\Omega\text{m}^2$ at 4.2 K where $A$ is the cross section
area. The renormalized resistivity\cite{Valet:prb93} is
$\rho^*_\text{Co}\sim75~\text{n}\Omega\text{m}$, and
$l^\text{Co}_\text{sf}=59~\text{nm}$ at 77 K.\cite{Piraux:epjb98} Thus
we can conclude that for Co $R_\text{F/N}/R_\text{sf}^\text{F}\approx 0.1$ as a
least estimate since the spin diffusion length should be longer at 4.2
K. This means that it is a valid approximation to disregard the
interface resistances for long enough samples. The N/S interface
resistance between Cu and Nb above the critical temperature is found
to be larger than the F/N resistance,\cite{Park:prb00}
$AR_\text{Cu/Nb}\sim 1.10~\text{f}\Omega\text{m}^2$, and would give
$R_\text{N/S}/R_\text{sf}^\text{F}\approx 0.2$. With Al as the superconducting
layer we expect no higher interface resistance. We may also argue that
the bulk resistance for dirty Cu/Co layers scales as
$AR_\text{bulk}^\text{Co}\sim 0.1L$[nm] f$\Omega$m$^2$,\cite{Kovalev:prb02}
where $L$ is the length of the layers expressed in nm. Thus the bulk
resistance for a slice of length $l_\text{sf}$ should be much larger
than the interface resistance. Note that in direct F/S interfaces the
reported interface resistance is considerably higher
$AR_\text{Nb/Co}\sim 3~\text{f}\Omega\text{m}^2$.\cite{Park:prb00}

The estimates above show that the interface resistances are much
smaller than the relevant bulk resistances with the materials chosen
here. We will later check that the change in resistance from normal to
superconducting state is larger than the interface resistances.  A
possible approximation is therefore to neglect the interface
resistances.  This allows us to effectively do calculations for an
F/S/F system with the boundary condition that the generalized
diffusive current should be continuous which implies that the function
$\theta$ is continuous at the interface. For strong ferromagnets the
superconducting proximity effect into the ferromagnet is negligible
and we have by continuity $\theta\to 0$ in the superconductor close to
the F/S interfaces. This means e.g.  that the gap vanishes at the
interface.  In this case, it is the bulk transport properties that
dominate the system, and there are no free parameters so that it is
possible to give an unambiguous description of the transport
properties. This is our aim in the rest of the paper.

The F/S/F system as shown in Fig. \ref{fig:pillar} was motivated by
the experiments of Gu {\it et al.}. However, in those experiments Py
was used for the ferromagnet, and because of the very short spin
diffusion length in this alloy ($l^\text{Py}_\text{sf}=5.5~\text{nm}$)
the interface resistances are of the same order as $R_\text{sf}^\text{F}$.
Consequently, in these experiments both the interface resistances and
the bulk resistance of Al are governed by superconductivity. Thus the
model discussed above is not applicable, and the resistance of the
spin polarizing interface must be taken into account. To be specific,
we no longer have that $\theta\to 0$ at the interfaces, and
superconductivity is not completely suppressed at the interface as in
the Co/Al/Co system.  Using the approximations discussed above in
calculations for the Py/Nb/Py system of Ref. \onlinecite{Gu:prb02}
would therefore give a too low $T_c$. Numerical simulations and
comparison with Ref.  \onlinecite{Gu:prb02} show that this is indeed
the case (not shown). A complete description of this experiment
requires boundary conditions for the spin polarizing interfaces given
by scattering theory. This would describe the proximity effect in N as
well as a reduction of the superconducting pairing amplitude close to
the interface. However, as noted by Huertas-Hernando {\it et
  al.},\cite{Huertas-Hernando:cond-mat02} this approach would require
full knowledge about the interface scattering matrix, which is
generally not available except for in simplified models at this
moment.

\section{\label{sec:numerics}Calculations}
%******************************************************************
We have performed numerical calculations for a Co/Cu/Al/Cu/Co spin
valve.  Parameters for the superconductor are mostly taken from Ref.
\onlinecite{Boogaard:prb04}. The bulk value of the pairing potential
at zero temperature is $\Delta_0=192~\mu e\text{V}$ and the critical
temperature $T_c=1.26~\text{K}$ with interaction parameter
$N_0\lambda/2=0.18$.\cite{Ibach:95} The normal state diffusion
coefficient of Al is $D=160~\text{cm}^2/\text{s}$, and the density of
states at Fermi level $N_0=2.2\cdot
10^{47}~\text{J}^{-1}\text{m}^{-3}$ corresponding to a resistivity
$\rho_\text{Al}^\text{N}=11~\text{n}\Omega\text{m}$. The normal
state spin-flip relaxation length by spin-orbit interaction is given
by the sample independent parameter $\varepsilon=l_\text{sf}/l\approx
30$,\cite{Jedema:prb03} and we assume that the elastic mean free path
is $l=37~\text{nm}$. This gives $l_\text{sf}=1.11~\mu\text{m}$ for
spin-orbit induced spin-flip. In calculations for magnetic impurity
induced spin-flip we take the normal state value of the spin-flip
length identical to the spin-orbit induced $l_\text{sf}$, but in
general this length is determined by the impurity concentration which
is sample specific. We take the length of the (identical)
ferromagnetic elements to be 100 nm with a bulk spin asymmetry
$\beta=0.4$.\cite{Bass:jmmm99}
\begin{figure}[htbp]
  \includegraphics[scale=0.67]{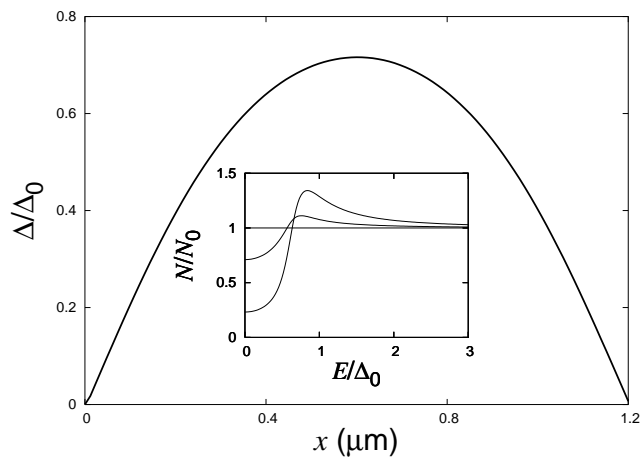}  
  \caption{Spatial variation of the pairing potential at $T/T_c=0.40$
    scaled by $\Delta_0$. {\it Inset}: Density of states at positions
    0.0, 0.2 and 0.6 $\mu$m into the S wire. The curve evaluated at
    $x=0.0~\mu$m is identical to the normal state DOS (flat curve).}
  \label{fig:gap}
\end{figure}
Figure \ref{fig:gap} shows the spatial variation of the pairing
potential resulting from complete suppression of superconductivity at
the F/S interfaces at reduced temperature $T/T_c=0.40$ for a 1.2
$\mu$m Al wire with magnetic impurities. The density of states at
various locations in the superconductor is shown in the inset, and
resembles the bulk BCS shape close to the center of the wire where the
gap is largest.

\begin{figure}[htbp]
  \includegraphics[scale=0.67]{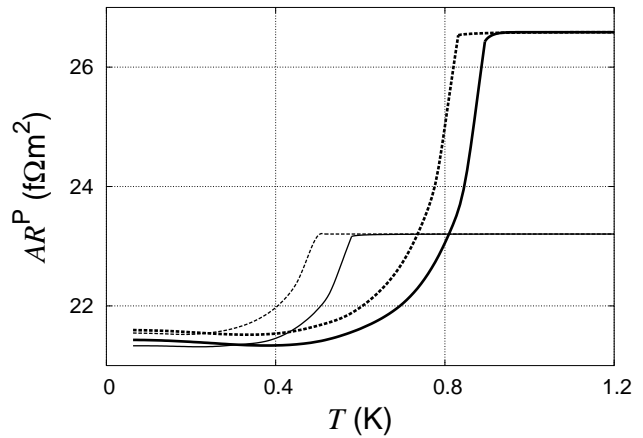}
  \caption{Temperature dependence of resistance in parallel
    geometry for spin-orbit (solid lines) and magnetic impurity
    (dashed lines) induced spin flip. The thick curves are with Al
    length 1200 nm, and the thin curves with Al length 900 nm. $T_c$
    is lowered by the presence of the magnetic impurities.}
  \label{fig:rparallel}
\end{figure}
A calculation of the resistance of the F/S/F system for parallel
magnetizations is shown in Fig. \ref{fig:rparallel}.  The
$AR^\text{P}$ values above the critical temperature agrees with
analytical results based on the Valet-Fert theory. Below $T_c$ the
resistance drops rapidly, but remains finite in the limit $T\to 0$.
The change in resistance from normal to superconducting state is of
the order of 2 - 6 f$\Omega$m$^2$ depending on the length of the
superconductor, and this change is larger than the typical interface
resistance, which should be checked as noted in Section
\ref{sec:model}. The resistance of the system below $T_c$ is due to
the F elements as well as the regions in the S wire next to the F/S
interface where there is conversion of current into
supercurrent.\cite{Boogaard:prb04} The systems with magnetic
impurities has the higher resistance as $T\to 0$ since the length of
the resistive region near the interfaces is longer. This is because
the conversion of current into supercurrent happens over a length
scale determined by the coherence length $\xi=\sqrt{\hbar D/2\pi
  \Delta}$ which for a superconductor with magnetic impurities is
longer since $\Delta$ is suppressed due to a term in the Usadel
equation \eqref{eq:retardedusadel}.

\begin{figure}[htbp]
  \includegraphics[scale=0.67]{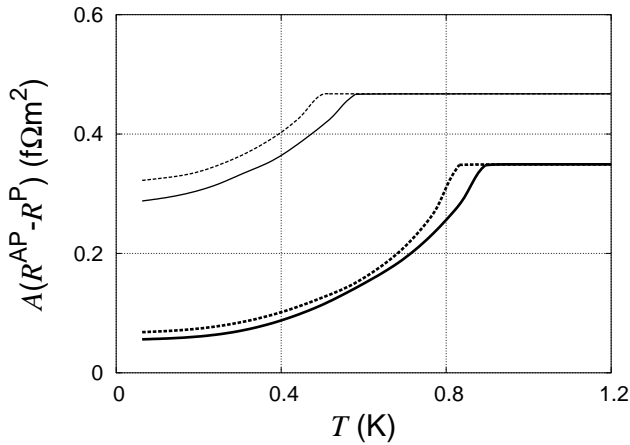}
  \caption{Temperature dependence of the magnetoresistance for
    spin-orbit (solid lines) and magnetic impurity (dashed lines)
    induced spin flip. The top set of curves is a system with Al
    length 900 nm, and the lower curves with Al length 1200 nm.}
  \label{fig:deltar}
\end{figure}
The dependence of the resistance on the magnetization configuration is
shown in Fig. \ref{fig:deltar} where the excess resistance $\Delta
R=A(R^\text{AP}-R^\text{P})$ is plotted as function of
temperature. We show curves for systems with only spin flip scattering
from magnetic impurities or spin-orbit interaction. The systems with
magnetic impurities provide a weaker suppression of the spin signal
than systems with spin-orbit interaction. The opposite could be
expected since as noted above
$\alpha^\text{m}_\text{TSTS}>\alpha^\text{so}_\text{TSTS}$.
However, the pairing potential is lower in a superconductor with
magnetic impurities due to the detrimental effect of the impurities on
superconductivity, and this is the dominant effect. This is confirmed
by simulations of systems with equal strengths of the pairing
potential, in which magnetic impurities is the strongest spin
relaxation mechanism. From Fig. \ref{fig:deltar} we see that the
difference in suppression of spin signal between spin-orbit and
magnetic impurity induced spin-flip is smaller for the longer
wires, since in this case the difference in $\Delta$ is also smaller.
For long wires the excess resistance tends to zero at low temperatures
as expected, because in this case the transport of spins through the
superconductor is completely suppressed.

\begin{figure}[htbp]
  \includegraphics[scale=0.67]{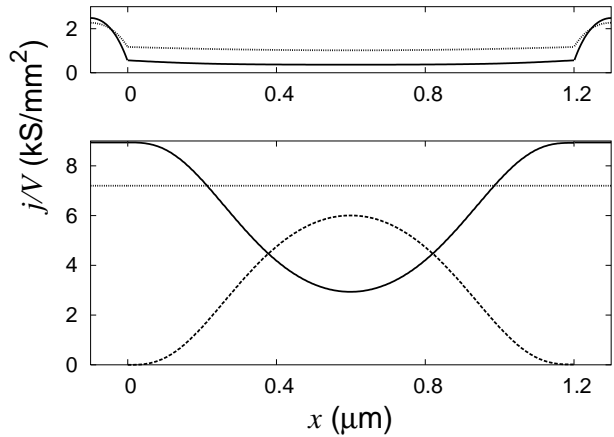}
  \caption{{\it Top panel}: Spatial dependence of the spin current for
    the F/S/F structure with Al length 1.2 $\mu$m at $T/T_c=0.40$
    (spin-orbit induced spin-flip). The normal state ($T>T_c$)
    spin-current is shown with dotted line. The F/S interfaces are at
    $x=0~\mu$m and $x=1.2~\mu$m. {\it Bottom panel}: Spatial
    dependence of the quasiparticle charge current (solid line) and
    supercurrent (dashed line) for the same system as in top panel.
    Normal state current is shown with dotted line.}
  \label{fig:icomb}
\end{figure}

\begin{figure}[htbp]
  \includegraphics[scale=0.67]{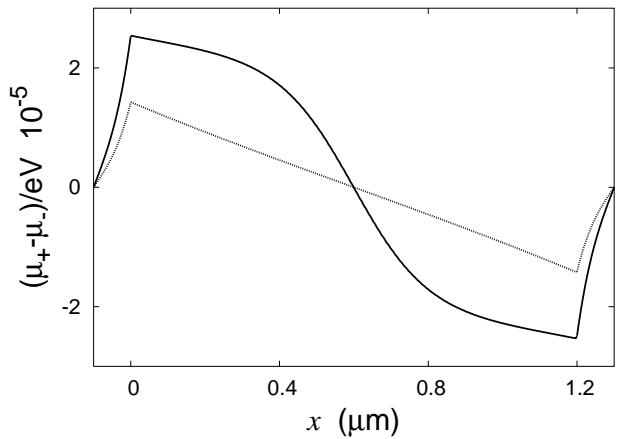}
  \caption{The spatial dependence of the spin potential for
    the F/S/F structure with Al length 1.2 $\mu$m at $T/T_c=0.40$
    (solid line) and $T>T_c$ (dotted line) (spin-orbit induced
    spin-flip). The F/S interfaces are at $x=0~\mu$m and
    $x=1.2~\mu$m.}
  \label{fig:mu}
\end{figure}

In Fig. \ref{fig:icomb} we show the spatial variation of the
quasiparticle charge and spin current and supercurrent for the
F/N/S/N/F spin valve with parallel magnetizations.
The charge current is constant in the F parts of the wire, and is
gradually converted into supercurrent in S. Spin current injection
into S is suppressed, as a comparison with the magnitude of spin
current in the normal state shows. This leads to spin-accumulation in
F at the interfaces. We see that the spin current is reduced below
$T_c$ inside the superconductor due to Cooper pairing. On the other
hand, the total charge current increases below $T_c$ due to the
reduced resistance of the superconductor. In Fig. \ref{fig:mu} we show
the spin accumulation $\mu_{\uparrow}-\mu_\downarrow$ for the same
system.
Comparison with the normal state shows that the spin accumulation is
larger in the S case, due to the reduced penetration of spins into S
and since the net spin current out of the reservoirs is larger in the
S case because the total resistance is lower. The spin accumulations
that build up at the interfaces are relaxed through spin flip in
S. These spin accumulations can be measured e.g. by tunnel coupling
between the superconductor and a third probe ferromagnet.

Qualitatively, our results for the MR are in agreement with the
experiment by Gu {\it et al.}\cite{Gu:prb02} A contribution from the
interfaces which is most probably important in the experiment, will
not qualitatively change the properties of the system except for a
higher $T_c$ as noted above.
Therefore, quantitative differences between the experiment and our
calculations using material parameters corresponding to the system in
Ref. \onlinecite{Gu:prb02} are not surprising. A more detailed
theoretical analysis, which accounts for interface resistance, should
be made to enable a quantitative comparison with the experiments of
Ref.~\onlinecite{Gu:prb02}, but this is beyond the scope of our
present work. We emphasize again, that our predictions are, however,
experimentally testable in Co/Al/Co spin valves, which can be
fabricated using state-of-the-art technology.

\section{\label{sec:conclusion}Summary and Conclusions}
%*******************************************************************
In conclusion we have studied spin transport properties of an F/S/F
trilayer. We have developed transport equations using the
quasiclassical theory of superconductivity and included the effects of
spin-flip relaxation. An experimental system is proposed where
interface resistance can be neglected and a simple description of the
physics is possible. For this system we have performed numerical
calculations of the magnetization-configuration dependent resistance.
This demonstrates the dependence of the spin transport suppression on
different spin flip mechanisms, i.e. magnetic impurities and
spin-orbit interaction.

\begin{acknowledgments}
  This work was supported in part by The Research Council of Norway,
  NANOMAT Grants No. 158518/431 and 158547/431, RTN Spintronics, the
  Swiss NSF, the NCCR Nanoscience, and EU via project
  NMP2-CT-2003-505587 'SFINx'
\end{acknowledgments}

\bibliography{fs}

\end{document}